\documentclass[letterpaper, 10 pt, conference]{ieeeconf}  

\IEEEoverridecommandlockouts                              

\usepackage{generic}
\usepackage{amsmath,amssymb,amsfonts}
\usepackage{graphicx}
\usepackage{textcomp}
\usepackage[english]{babel}
\usepackage[utf8x]{inputenc}
\usepackage[colorinlistoftodos]{todonotes}
\usepackage{comment}
\usepackage{algorithm}
\usepackage[noend]{algpseudocode}
\usepackage[normalem]{ulem}
\usepackage{balance}
\renewcommand{\baselinestretch}{0.98}
\long\def\comment#1{}
\usepackage[hidelinks]{hyperref}

\usepackage{graphicx}      
\usepackage{amsmath}
\usepackage{amssymb}
\usepackage{algorithm}
\usepackage[noend]{algpseudocode}
\newtheorem{theorem}{Theorem}

\usepackage{xcolor}
\usepackage{caption}
\usepackage{subcaption}
\usepackage[scr=boondoxo, scrscaled=.98]{mathalfa} 
\usepackage{setspace}
\usepackage{soul}
\usepackage{multirow}
\usepackage{hhline}
\usepackage{booktabs}
\usepackage{float}
\usepackage{mathtools}
\usepackage{hyperref}
\newtheorem{corollary}[theorem]{Corollary}
\newtheorem{definition}{Definition}

\newtheorem{proposition}[theorem]{Proposition}

\newtheorem{remark}{Remark}
\newtheorem{standingassumption}{Standing Assumption}

\newcommand{\RR}{\mathbb{R}}
\newcommand{\CC}{\mathbb{C}}

\newcommand{\LL}{\mathcal{L}}

\newcommand{\norm}[1]{\left\lVert#1\right\rVert}

\DeclareMathOperator*{\arginf}{arg\,inf}

\usepackage{generic}
\usepackage{cite}
\usepackage{textcomp}
\usepackage{soul}
\def\BibTeX{{\rm B\kern-.05em{\sc i\kern-.025em b}\kern-.08em
    T\kern-.1667em\lower.7ex\hbox{E}\kern-.125emX}}

\newif\ifpreprint
\preprinttrue

\title{\LARGE \bf
Low-dimensional observer design for stable linear systems \\ by model reduction  
}

\author{M.F. Shakib$^{1}$, M. Khalil$^{2,3}$, and R. Postoyan$^{2}$ 
\thanks{$^{1}$Department of Electrical and Electronic Engineering, Imperial College London, London, UK {\tt (m.shakib@imperial.ac.uk)}.}
\thanks{$^{2}$Université de Lorraine, CNRS, CRAN, F-54000 Nancy, France {\tt{(firstname.surname@univ-lorraine.fr).}}}
\thanks{$^{3}$Universit{\' e} de Lorraine, GREEN, F-54000 Nancy, France.}
\thanks{This work has been partially supported by the Engineering and Physical Sciences Research Council, Grant EP/X033546/1.}
}

\begin{document}

\maketitle
\thispagestyle{empty}
\pagestyle{empty}

\begin{abstract}
This paper presents a low-dimensional observer design for stable, single-input single-output, continuous-time linear time-invariant (LTI) systems.
Leveraging the model reduction by moment matching technique, we approximate the system with a reduced-order model.
Based on this reduced-order model, we design a low-dimensional observer that estimates the states of the original system. 
We show that this observer establishes exact asymptotic state reconstruction for a given class of inputs tied to the observer's dimension.
Furthermore, we establish an exponential input-to-state stability property for generic inputs, ensuring a bounded estimation error. Numerical simulations confirm the effectiveness of the approach for a benchmark model reduction problem.
\end{abstract}

\section{Introduction}\label{sect:introduction}

In many applications, it is crucial to know the state of the dynamical system at hand, e.g., for state feedback or monitoring purposes.
However, very often only the system output is measured rather than the full state. An observer-based approach may then be envisioned to estimate the state from input-output data. Such an observer takes the form of an auxiliary dynamical system, whose dimension is typically equal to or larger than the system dimension \cite{bernard-et-al-arc22(observer)}.

When the state dimension of the system is excessively large, designing as well as implementing an observer in real-time on devices with limited resources might be challenging or even infeasible. This challenge may arise when the plant model is obtained by spatially discretizing partial differential equations, like in, e.g., \cite{Dey_et_al_ACC_2014,blondel-et-al-tcst2018,khalil-et-al-tcst24}, or when considering networked systems, like in \cite{khalil-et-al-cdc2024}, for instance. 
A common technique to obtain an observer to reconstruct the full state of the system with a dimension smaller than that of the plant model is to proceed with a so-called \emph{reduced-order observer}  \cite{darouach2002reduced,Astolfi_Karag_Ortega_08,Arcak_Kokotovic-Aut01}. 
Reduced-order observers only estimate the state that cannot be directly inferred from measurements. 
By doing so, the observer dimension is reduced by the number of measured outputs, which is typically small. 
A different approach is proposed in  \cite{sadamoto-et-al-cdc2013}, where it is explained how model reduction techniques can be exploited to derive a low-dimensional observer with guaranteed performance from a given Luenberger observer for an LTI plant model.  When the plant exhibits a network structure, an alternative approach consists in designing a distributed observer, see, e.g., \cite{han2018simple,wang2017distributed}, or an average state observer \cite{sadamoto-et-al-tcn2016}. Ad-hoc solutions are also available for specific applications, like in, e.g., \cite{khalil-et-al-cdc2024,plett2009efficient} when dealing with battery packs. Despite these achievements, we generally lack methodological tools to design low-dimensional observers, which are able to generate estimates of the full state vector.

\begin{figure}
    \centering
    \includegraphics[width= 0.48\textwidth]{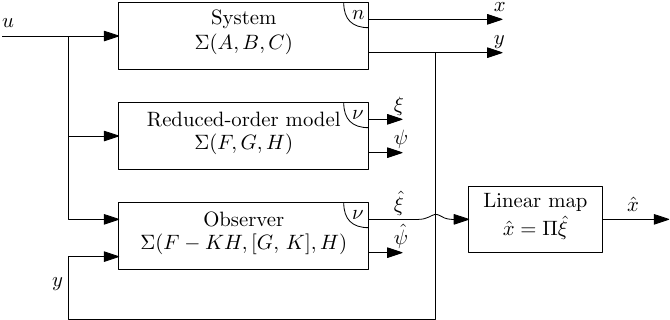}
    \caption{\label{fig:Schematic_observer}
    Schematic overview of the proposed observer scheme with the system state dimension $n$ and the observer state dimension $\nu $, typically smaller than $n$. The observer is designed based on a reduced-order model of dimension $\nu$.}
    \vspace{-5mm}
\end{figure}

In this context, the problem considered in this article is to design an observer that reconstructs the full state of a (possibly large-scale) stable single-input single-output linear time-invariant (LTI) system, where the dimension of the observer dynamics is freely selected by the user as explained next. The main originality is to exploit recent advances in model order reduction  \cite{astolfi2010model} for this purpose. The approach that we envision is, in the first step, to approximate the dynamics of the system by a reduced-order model. 
Specifically, we use the interconnection-based moment matching technique~\cite{astolfi2010model}. 
A distinctive feature of this technique is that it provides a \emph{match} between the steady-state response of the original plant and the steady-state response of the reduced-order model for user-defined classes of inputs characterized by so-called interpolation points.
The dimension of the reduced-order model then directly relates to the number of interpolation points and, thereby, to the considered class of inputs.
In the second step, we design an observer for the reduced-order model that has the same state dimension as the reduced-order model. The observer dimension is thus directly related to the user-defined class of inputs considered when performing the model order reduction. 
Finally, in the third step, we map the reconstructed state of the reduced-order model back to the non-reduced state of the original system.
By doing so, we obtain an observer, which is able to reconstruct the full state of the original plant model, while typically having a much lower dimension. 
Moreover, the observer design 
only involves the matrices of the reduced model thereby facilitating its use as well as its implementation.   
Figure~\ref{fig:Schematic_observer} provides a schematic overview of the proposed approach.

We establish that the estimation error, i.e., the mismatch between the plant's full state and its estimate generated by the low-dimensional observer, satisfies a global exponential input-to-state stability property. The  ``disturbance'' term in this stability property corresponds to the mismatch between the actual system input
and the closest input signal belonging to the class considered in the model reduction step through the choice of the interpolation points.
Particularly, for inputs corresponding to these interpolation points, we show that this ``disturbance term'' is zero, and thus the estimation error tends to zero exponentially since the moment matching model reduction technique is exact for these inputs.
This is despite a possibly, significantly smaller state dimension of the implemented observer, compared to the state dimension of the original system. In contrast to \cite{sadamoto-et-al-cdc2013}, we perform the model reduction directly on the plant rather than on a previously designed observer for the original system. This simplifies the design process, as we rely solely on the reduced-order model to synthesize the low-dimensional observer. Furthermore, exact asymptotic estimation is guaranteed whenever the input applied to the plant belongs to the class of inputs used for model reduction.

As mentioned above, we concentrate on stable single-input single-output linear time-invariant systems as a first step towards the exploitation of model order reduction techniques for observer design. 
As follow-up studies, we envision extending the proposed approach to different system classes such as, for example, unstable and nonlinear systems.

The remainder of this paper is organized as follows. 
Section \ref{sect:problem-statement} introduces the considered problem. 
Section \ref{sect:preliminaries-model-reduction} provides the  background material on model reduction that is exploited in this work to design low-dimensional observers.  
Section \ref{sec:mainresult} presents the main theoretical contributions.  
Section \ref{sec:example} provides numerical examples showcasing the performance of the proposed method on a high-dimensional system. 
Section \ref{sect:conclusions} concludes the paper with a discussion of key findings and potential future directions.
\ifpreprint{}
\else{
The proofs are omitted for space reasons and can be found in~\cite{shakib2023low}.}
\fi
\\

\noindent\textbf{Notation:} Let $\mathbb{R}$, $\mathbb{R}_{\geq 0}$, and $\mathbb{R}_{> 0}$ denote the set of real, non-negative real, and positive real numbers, respectively. Let $\mathbb{N}$ and $\mathbb{N}_{>0}$ be the set of non-negative and positive integers, respectively. We use $\mathbb{C}$, $\mathbb{C}^{0},\CC^-$, and $\mathbb{C}^{+}$ to denote the set of complex numbers, and the set of complex numbers with zero real part, with negative real part and with positive real part, respectively. Given $n\in\mathbb{N}_{>0}$,  $\mathcal{L}_{\infty,n}$ stands for the set of functions from $\mathbb{R}_{\geq 0}$ to $\mathbb{R}^n$ that are locally essentially bounded and measurable, and we omit the index $n$ when the dimension is clear from the context. Given $f\in\mathcal{L}_\infty$ and $t\in\mathbb{R}_{\geq 0}$, $\|f\|_t$ stands for $\text{ess\,}\sup_{t'\in[0,t]}\|f(t')\|$ where $\|\cdot\|$ stands for the Euclidean norm. We use the notation $\langle \cdot,\cdot\rangle$ for the scalar product of the considered Euclidean space, and $\nabla$ for the nabla operator. Given a real, symmetric matrix $P$, its largest
and smallest eigenvalues are denoted by $\lambda_\text{max}(P)$ and
$\lambda_\text{min}(P)$, respectively. We write rank$(A)$ to denote the rank of a real matrix $A$. Given a square real matrix $A$, $\sigma(A)$ is the spectrum of $A$ and $A\succ0$ means that $A$ is symmetric, positive definite. 
Furthermore, $\text{blockdiag}(A_1,A_2)$ denotes a block-diagonal matrix of square submatrices $A_1$ and $A_2$. We write $||A||$ to denote the $2$-induced norm of the real square matrix $A$. 
The imaginary unit $\sqrt{-1}$ is denoted by $j$.

\section{Problem statement}\label{sect:problem-statement}
\label{S:PS}
We consider single-input single-output LTI systems described by the next state space representation
\begin{equation}\label{eq:sys_LTI}
    \dot{x} = Ax + Bu, \quad 
    y = Cx,
\end{equation}
where, at time $t\in\mathbb{R}_{\geq 0}$, $x(t) \in \mathbb{R}^{n}$ is the state vector with $n\in\mathbb{N}_{>0}$, $u(t) \in \mathbb{R}$ is the input, and $y(t)\in\mathbb{R}$ is the output. 
The matrices $A, B, \text{ and } C$ are real-valued and have appropriate dimensions. 
We assume that the signal corresponding to $u$ belongs to $\mathcal{L}_\infty$. 

In many applications, the state $x$ of the system has to be known, while only the output $y$ of the system can be measured.
In such a case, an observer, for example, a Luenberger observer, can be designed to reconstruct the unknown state  
when the pair $(A,C)$ is detectable.
In particular, for system~\eqref{eq:sys_LTI}, such a Luenberger observer takes the  form
\begin{align}\label{eq:observer_full}
    \dot{\hat{x}} = A \hat x + Bu + M(y-\hat y), \quad 
    \hat y = C\hat x,
\end{align}
where, at time $t\in\RR_{\ge0}$, $\hat x(t) \in \mathbb{R}^{n}$ is the estimated state vector,
 $\hat y(t)\in\mathbb{R}$ is the estimated output, and  $M\in\RR^{n}$ is the observer gain. 
Assuming $(A,C)$ is detectable, a gain $M$ can always be designed such that $A-MC$ is Hurwitz, i.e., $\sigma(A-MC)\subset \mathbb{C}^-$,
which results in the global exponential convergence of the error $e:=x-\hat x$ to zero  along the solutions to (\ref{eq:sys_LTI}), (\ref{eq:observer_full}), see \cite{Hespanha2018}.

As evident from observer~\eqref{eq:observer_full}, the dimension of $\hat x$ is $n$, which is equal to the state dimension of the original system~\eqref{eq:sys_LTI}.
For large-scale systems, this presents a substantial challenge not just in designing the observer gain~$M$, but also in implementing the observer in real-time on devices with limited resources. The objective of this work is to provide a method to design a state-observer for system (\ref{eq:sys_LTI}), whose dimension can be freely tuned by the user depending on the considered class of signals to which $u$ belongs to. Moreover, this observer has to generate estimates of the original full state vector $x$, and not only of the unmeasured states as in reduced-order observers. We propose to leverage the model order reduction techniques in \cite{astolfi2010model} for this purpose. The next section reviews relevant background.
Throughout this article, we pose the following standing assumption.
\begin{standingassumption}[SA1]
    System~\eqref{eq:sys_LTI} is minimal and the matrix $A$ in~\eqref{eq:sys_LTI} is Hurwitz, i.e., $\sigma(A) \subset \CC^-$. \hfill $\Box$
\end{standingassumption}

\begin{remark} By SA1, taking $M=0$ in (\ref{eq:observer_full}) leads to $A-MC$ Hurwitz. Still, the implementation of (\ref{eq:observer_full}) may be challenging when $n$ is large. In addition, taking $M=0$ may not be a desired choice as the convergence and robustness properties of the observer are then limited by those of the plant (\ref{eq:sys_LTI}).
\hfill $\Box$
\end{remark}

\section{Preliminaries on model reduction\\by moment matching}\label{sect:preliminaries-model-reduction}
\label{S:MM}

The Luenberger observer introduced in~\eqref{eq:observer_full} reconstructs the full state for any input $u \in \LL_{\infty}$.
However, for specific classes of inputs, as formalized in Section~\ref{S:input_class}, the observer dimension can be reduced based on model reduction by moment matching recalled in Section~\ref{S:mm}.

\subsection{Inputs generated by signal generators}\label{S:input_class}
We consider inputs that are generated by so-called signal generators described by the equations
\begin{equation}\label{eq:SG}
    \dot \omega = S\omega, \quad u = L\omega,
\end{equation}
where $\nu\in\mathbb{N}_{>0}$ with $\nu<n$, and, at time $t\in\RR_{\ge 0}$, $\omega(t)\in\RR^{\nu}$ is its state and $u(t)\in\RR$ is the generated input.
The specific choice of $\nu$ and of the matrices $S\in\RR^{\nu\times\nu}$ and $L \in \RR^{1\times \nu}$ defines a class of inputs that can be generated by~\eqref{eq:SG}.

Let $S\in\RR^{\nu\times\nu}$ be such that $\sigma(S)\subset\CC^0$ and $\sigma(S)$ simple.
Then, the eigenvalues of $S$ are located on the imaginary axis and can be interpreted as pairs of complex conjugate interpolation frequencies.
For example, the case~$S=0$ (hence~$\nu=1$), 
system (\ref{eq:SG}) can generate \emph{any} constant input~$u$. 
Similarly, the case~$\nu =2$ and~$\sigma(S) =\{-jf,jf\}$ for some~$f\in\RR_{>0}$, results in inputs~$u(t) = A_s\sin(\pm ft+\varphi)$ for some~$A_s\in\RR$ and~$\varphi\in\RR$, depending on the initial condition of the generator. In this case, the system (\ref{eq:SG}) can generate any sine wave with frequency $f$. 

As we will see, the proposed low-dimensional observer achieves asymptotically zero reconstruction error for any input generated by the signal generator~\eqref{eq:SG}.
Therefore, a larger~$\nu$, i.e., a larger number of interpolation frequencies, provides a zero asymptotic estimation error for a larger class of inputs, e.g., sine waves with different frequencies.
As the dimension~$\nu$ is a user choice, the dimension of the observer can thus be made arbitrarily small.
The particular choice of the interpolation frequencies is  problem-specific.

We emphasize that the results of this article are not limited to only input generated by the signal generator~\eqref{eq:SG}.
For inputs that cannot be generated by~\eqref{eq:SG}, we show that the reconstruction error is generally non-zero, but bounded.
Notably, this bound depends on the mismatch between the actual input and the closest one that can be generated by the signal generator~\eqref{eq:SG}.
We formalize this result in Section~\ref{sec:mainresult}.

In the next section, we recall a family of reduced-order models that achieve so-called moment matching.
This family is used to design the low-dimensional observer in Section~\ref{sec:mainresult}.
We summarize the required standing assumption on the signal generator~\eqref{eq:SG} first.
\begin{standingassumption}[SA2]\label{as:SG_standard}
    The pair $(S,L)$ in~\eqref{eq:SG} is observable and satisfies $\sigma(S) \cap \sigma(A) = \emptyset$. \hfill $\Box$
\end{standingassumption}

\subsection{Model reduction by moment matching}\label{S:mm}
Given the signal generator (\ref{eq:SG}), we consider the reduced-order model of the form
\begin{equation}\label{eq:ROM}
    \dot \xi = F \xi + Gu, \quad \psi = H \xi,
\end{equation}
for system~\eqref{eq:sys_LTI}, where, at time $t\in\RR_{\ge0}$, $\xi(t)\in \RR^\nu$ is the reduced-order state with $\nu$ as in (\ref{eq:SG}) and $\psi(t) \in \RR$ is the reduced-order model output. 
The matrices $F,G,$ and $H$ are real-valued and have appropriate dimensions.
In this article, we employ the model reduction by moment matching technique. 
Let us first recall the notion of moment and interpolation points.

\begin{definition}[\!\!\cite{astolfi2010model}]\label{def.moments}
Consider matrices $S \in \RR^{\nu\times\nu}$, $L \in \RR^{1\times \nu}$, and let 
$\Pi\in\RR^{n\times\nu}$ satisfy the Sylvester equation
\begin{equation}\label{eq:Sylvester_eqn}
    A\Pi + BL = \Pi S.
\end{equation}
Then, the matrix $C\Pi$ is called the \emph{moment} of system~\eqref{eq:sys_LTI} at~$\sigma(S)$, where $\sigma(S)$ is the set of \emph{interpolation points}.
\hfill$\Box$
\end{definition}

By SA1 and SA2, the solution~$\Pi\in\RR^{n\times\nu}$ in~\eqref{eq:Sylvester_eqn} is guaranteed to exist and to be unique. Furthermore, 
rank$(\Pi)=\nu$.
Indeed, uniqueness follows from~$\sigma(S)\cap\sigma(A) = \emptyset$, implied by SA2, and the rank of $\Pi$ is equal to $\nu$ by  the controllability of $(A,B)$, implied by SA1, and the observability of $(S,L)$, implied by SA2, see~\cite{de1981controllability} for details.
In Definition~\ref{def.moments}, the matrices $S$ and $L$ are interpreted as the matrices defining the signal generator~\eqref{eq:SG}, thus playing a key role in the moment matching problem.
Moment matching aims at finding appropriate matrices $F,G,$ and $H$  such that the reduced-order model~\eqref{eq:ROM} shares the same moment as the original system~\eqref{eq:sys_LTI} at the user-defined set~$\sigma(S)$, as formalized next.

\begin{definition}
    Consider matrices $S \in \RR^{\nu\times\nu}$, $L \in \RR^{1\times \nu}$, $F\in\RR^{\nu\times\nu}$, $G\in\RR^{\nu}$, and $H\in\RR^{1\times\nu}$ such that $\sigma(S)\cap\sigma(F) = \emptyset$.
    Let $C\Pi$ be the moment of system~\eqref{eq:sys_LTI} at $\sigma(S)$, where $\Pi\in\RR^{n\times \nu}$ is the unique solution to~\eqref{eq:Sylvester_eqn}.
    Let $HP$ be the moment of model~\eqref{eq:ROM} at $\sigma(S)$, where $P \in \RR^{\nu\times\nu}$ is the unique solution to $FP + GL = PS$. Then, model~\eqref{eq:ROM} is said to achieve \emph{moment matching} if $HP = C\Pi$. 
    \hfill$\Box$
\end{definition}

Given a pair of matrices $(S,L)$ and taking $P=I$, \cite{astolfi2010model} proposed a family of reduced-order models (\ref{eq:ROM}) 
that achieve moment matching at~$\sigma(S)$, as recalled below. 

\begin{theorem}[\!\!\cite{astolfi2010model}]\label{thm:Family}
Consider system~\eqref{eq:sys_LTI} and matrices $S \in \RR^{\nu\times\nu}$ and $L \in \RR^{1\times \nu}$.
Then, for any $G \in \RR^{\nu}$ such that $\sigma(S)\cap \sigma(S-G L) = \emptyset$, the model~\eqref{eq:ROM} with $F=S-GL$ and $H=C\Pi$ with $\Pi \in \RR^{n \times \nu}$ the unique solution to \eqref{eq:Sylvester_eqn}, achieves moment matching at $(S,L)$.\hfill$\Box$
\end{theorem}

Theorem~\ref{thm:Family} characterizes all the models~\eqref{eq:ROM} of dimension~$\nu$ that achieve moment matching.
The family is parameterized by $G\in\RR^{\nu}$ with the mild condition $\sigma(S)\cap \sigma(S-G L) = \emptyset$, since by observability of the pair $(S,L)$, the elements of the set $\sigma(S-G L)$ can be located at any desired location different from $\sigma(S)$.
The approximation quality depends on the particular $G$ selected, for example, the one that minimizes the $\mathcal{H}_\infty$-norm error~\cite{shakib2021optimal_,Shakib2024optimal}.
We refer to~\cite{scarciotti2024interconnection} for a recent survey.
The family presented in Theorem~\ref{thm:Family} is called a family of reduced-order models whenever $\nu < n$.

It is noted that the reduced-order model~\eqref{eq:ROM} as well as the signal generator~\eqref{eq:SG} are fictitious in the problem at hand.
Therefore, when implementing the low-dimensional observer, neither the model~\eqref{eq:ROM} nor the signal generator~\eqref{eq:SG} need to be implemented.

\section{Main results}\label{sec:mainresult}
This section contains the main results of this article.
First, the proposed observer is introduced in Section~\ref{S:MR:A}.
After that, in Section~\ref{S:MR:convergence}, an analysis is given that shows the convergence of the observer estimates for inputs generated by the signal generator~\eqref{eq:SG} as well as for generic inputs.
Finally, conditions for the convergence of the observer are given in Section~\ref{S:MR:design}.

\subsection{Low-dimensional observer and a preliminary analysis}
\label{S:MR:A}
We propose the following observer based on~\eqref{eq:ROM}:
\begin{equation}\label{eq:observer}
    \dot{\hat{\xi}} = (S-GL)\hat{\xi} + Gu + K(y - \hat{\psi}), \quad 
    \hat{\psi} = C\Pi \hat{\xi},
\end{equation}
where $K\in\RR^{\nu}$ is the observer gain.
Note that the state dimension of this observer is $\nu (< n)$.
We emphasize that the system output~$y$ is injected rather than the reduced-order model output~$\psi$ in the observer.
Given $\hat \xi \in \RR^\nu$, we obtain the full state estimate $\hat x \in \RR^n$ for $x \in \RR^n$ through the relation
\begin{equation}
    \hat x = \Pi \hat\xi,
\end{equation}
with $\Pi$ the unique solution of~\eqref{eq:Sylvester_eqn}.

To analyze the convergence properties of the proposed observer, 
we introduce the estimation error
$e_{x\hat\xi} \coloneqq x - \Pi\hat \xi
        = (x - \Pi\omega ) - \Pi (\hat \xi - \omega).$
We first establish the dynamics of
$x$ to $\Pi\omega$, i.e., of $e_{x\omega}\coloneq x -\Pi\omega$. 
Taking the time derivative of $e_{x\omega}$ along the solutions to (\ref{eq:sys_LTI}) and using $\dot\omega=S\omega$ results in
    \begin{align}
    \label{eq:exw_dyn}
    \hspace{-1mm}
       \dot e_{x\omega} = \! \dot x  \! - \! \Pi\dot\omega
        = \! Ax  +  Bu \! -\! \Pi S\omega  = \!A e_{x\omega} + B(u\!-\!L\omega)
    \end{align}
where the Sylvester equation~\eqref{eq:Sylvester_eqn} has been used.
Next, we write the dynamics of
$\hat \xi$ to $\omega$, i.e., of $e_{\hat\xi\omega} \coloneqq \hat \xi - \omega$.
Again, taking its time derivative yields along the solutions to (\ref{eq:observer}) and using $\dot \omega = S\omega$ results in
\begin{align}\label{eq:exihatw_dyn}
\begin{split}
    \dot e_{\hat\xi\omega} &= \! (S-GL)\hat{\xi} + Gu + K(y - \hat{\psi}) - S \omega\\
        &= \!(S-GL-KC\Pi)e_{\hat\xi\omega} \! + \! KCe_{x\omega} \! + \! G(u-L\omega).
    \end{split}
\end{align}
In conclusion, the overall dynamics of $e_{x\omega}$ in~\eqref{eq:exw_dyn} and $e_{\hat\xi\omega}$ in~\eqref{eq:exihatw_dyn} can then be written as follows:
\begin{align}\label{eq:error_dyn}
    \underbrace{\begin{bmatrix}
        \dot e_{x\omega} \\ \dot e_{\hat\xi\omega}
    \end{bmatrix}}_{\dot e} &= 
    \underbrace{\begin{bmatrix}
        A & 0 \\ KC & S-GL-KC\Pi
    \end{bmatrix}}_{=:\Xi}
    \underbrace{\begin{bmatrix}
        e_{x\omega} \\ e_{\hat\xi\omega}
    \end{bmatrix}}_{=:e} +
    \underbrace{\begin{bmatrix}
        B \\ G
    \end{bmatrix}}_{=:\Psi} (u-L\omega) \notag \\
    &= \Xi e + \Psi(u-L\omega),\\
    e_{x\hat\xi} &= \underbrace{\begin{bmatrix}
        I & -\Pi
    \end{bmatrix}}_{=:\Phi}e = x - \Pi \hat \xi \notag ,
\end{align}
where the output $e_{x\hat\xi}$ is the estimation error that we are particularly interested in, namely, $x-\Pi \hat \xi$.

The matrix $\Xi$ has a block-diagonal structure, where the matrix $A$ is Hurwitz by SA1.
Hence, $\Xi$ is a Hurwitz matrix if and only if $S-GL-KC\Pi$ is also Hurwitz. As evident from~\eqref{eq:error_dyn}, the estimation error depends on the input $u-L\omega$.
\ifpreprint{
Viewing the system~\eqref{eq:error_dyn} as a multi-input system with independent inputs $u$ and $\omega$ and provided that the matrix $\Xi$ is Hurwitz, we can give an input-to-state stability result, whose proof is given in the appendix.
}
\else{
Viewing the system~\eqref{eq:error_dyn} as a multi-input system with independent inputs $u$ and $\omega$ and provided that the matrix $\Xi$ is Hurwitz, we can give an input-to-state stability result.
}
\fi

\begin{proposition}\label{proposition:observer_convergence_generic}
  Consider system \eqref{eq:error_dyn} and suppose that $\Xi$ is Hurwitz.
Then, there exist constants $c_1, c_2, c_3 \in \RR_{>0}$ such that, for any inputs $u \in \mathcal{L}_{\infty,1}$  and $\omega \in \mathcal{L}_{\infty,\nu}$, any corresponding solution $e$ to~\eqref{eq:error_dyn} satisfies, for all $t \geq 0$,
    \begin{multline}\label{eq:ISS_generic_inputs}
        \hspace{-0.3cm}\|e_{x\hat \xi}(t) \| \leq c_1 \left(\|e_{xw}(0)\| + \|e_{\hat \xi \omega} (0) \|\right)\exp{(-c_2t)}\\
        + c_3 \left\| u-L\omega \right\|_t.
    \end{multline}
    \hfill$\Box$
\end{proposition}

Hence, if $\Xi$ is Hurwitz, then there exists an upper-bound on the norm of the estimation error~$e_{x\hat \xi}$ that consists of two terms.
The first term, related to the exponential function, addresses the effect of the initial conditions and drops exponentially to zero as  $t$ increases.
The second term, related to the input $u-L\omega$, addresses the effect of external inputs.

In Proposition~\ref{proposition:observer_convergence_generic}, the inputs~$u$ and~$\omega$ are viewed as two external inputs.
However, note that~$\omega$ is not an input to  system~\eqref{eq:sys_LTI}, nor to the reduced-order model in \eqref{eq:ROM}, or the observer~\eqref{eq:observer}.
Therefore, in fact, the output $e_{x\hat\xi}$ of the error system~\eqref{eq:error_dyn} can only be interpreted as the estimation error if $\omega$ is a trajectory of the signal generator~\eqref{eq:SG}, starting from some initial condition~$\omega_0\in\RR^n$.
In this context, the bound in Proposition~\ref{proposition:observer_convergence_generic} holds for any trajectory generated by the signal generator~\eqref{eq:SG}, i.e., any \emph{initial condition} of the generator.
The next section presents a convergence analysis for the observer~\eqref{eq:observer} for general inputs~$u$ and in the case in which~$\omega$ is a solution of the signal generator~\eqref{eq:SG}.

\begin{remark}
    It is evident from~\eqref{eq:error_dyn} that the matrix $A$ must be Hurwitz for $\Xi$ to be Hurwitz.
    Future work will address systems with an $A$ matrix that is not Hurwitz.%
    \hfill $\Box$
\end{remark}

\subsection{Estimation error convergence guarantees}
\label{S:MR:convergence}

The bound in Proposition~\ref{proposition:observer_convergence_generic} is an input-to-state stability result that depends on 
$\left\| u-L\omega \right\|_t$ (see~\ref{eq:ISS_generic_inputs}), where $\omega$ can be any trajectory 
of the signal generator~\eqref{eq:SG}.
As one of our main results, we show next that the bound can be made tighter by letting $\omega$ be a trajectory of the signal generator~\eqref{eq:SG} with a specific initial~condition.

\begin{theorem}~\label{thm:observer_convergence_generic} 
    Consider system \eqref{eq:sys_LTI}, \eqref{eq:SG}, and \eqref{eq:observer} and let $\Pi$ be the unique solution of the Sylvester equation~\eqref{eq:Sylvester_eqn}.
    Suppose the following holds.
    \begin{enumerate}
    \item[(i)] $G$ is such that $\sigma(S)\cap\sigma(S-GL)=\emptyset$.
    \item[(ii)] $\Xi$ in~\eqref{eq:error_dyn} is Hurwitz.
    \end{enumerate}
    Then,
    for any input $u \in \mathcal{L}_{\infty,1}$, any corresponding solution $(x, \hat{\xi})$ to~\eqref{eq:sys_LTI}, \eqref{eq:observer} satisfies, for all $t \geq 0$,
    \begin{multline}\label{eq:ISS_generic_inputs_tight_bound}
        \|e_{x\hat \xi}(t) \| \leq \\
        c_1 \left(\|e_{x\omega}(0)\| + \|e_{\hat \xi \omega} (0) \|\right)\!\exp{(-c_2t)} + c_3 \tau(t),
    \end{multline}
    where $c_1$, $c_2$, $c_3\in\mathbb{R}_{>0}$ are as in 
    Proposition~\ref{proposition:observer_convergence_generic}, $e_{x\omega}(0) \coloneqq x(0)- \Pi\omega_0$, $e_{\hat \xi \omega}(0) \coloneqq \hat \xi(0) - \omega_0$, and
    \begin{equation}
        \tau(t) \coloneqq \norm{u(t')-L\exp{(St')\omega_0}}_t    
    \end{equation}
    with $\omega_0 \in \RR^\nu$ defined as
    \begin{equation}\label{eq:choice_w0}
        \omega_0 \in \arginf_{\omega_0' \in \RR^\nu}\text{ess.sup}_{0\le t'\le t} \norm{u(t')-L\exp{(St')}\omega_0'}.
        \vspace{-3mm}
    \end{equation}
    \hfill$\Box$
\end{theorem}

\ifpreprint{
\begin{proof}
    First, note that by SA2 and item (i) of Theorem~\ref{thm:observer_convergence_generic}, 
    the reduced-order model~\eqref{eq:ROM} achieves moment matching at $\sigma(S)$, see Theorem~\ref{thm:Family}.
    Hence, as shown in the analysis in Section~\ref{S:MR:A}, the  estimation error $e_{x\hat \xi}$ satisfies the dynamics~\eqref{eq:error_dyn}, where~$\omega$ is a trajectory of the signal generator~\eqref{eq:SG}.
    In addition, since the matrix $\Xi$ is assumed to be Hurwitz, the bound~\eqref{eq:ISS_generic_inputs} holds for generic inputs $u$ and~$\omega$.
    
    From here, note that $\omega(t) = \exp{(St)}\omega_0$ is the solution of the signal generator~\eqref{eq:SG}, starting from some initial condition~$\omega_0$.
    Substituting this into~\eqref{eq:ISS_generic_inputs} yields
    \begin{multline}
        \|e_{x\hat \xi}(t) \| \leq c_1 \left(\|e_{xw}(0)\| + \|e_{\hat \xi \omega} (0) \|\right)\exp{(-c_2t)}\\
        + c_3 \left\| u-L \exp{(St)\omega_0} \right\|_t.
    \end{multline}
    Since $\omega_0$ is free to choose, we can choose it as in~\eqref{eq:choice_w0}, which leads to the definition of $\tau$ in the theorem statement and to the bound~\eqref{eq:ISS_generic_inputs_tight_bound}.
    This completes the proof.
\end{proof}
}\fi

The bound~\eqref{eq:ISS_generic_inputs_tight_bound} still consists of two terms, like in Proposition \ref{proposition:observer_convergence_generic}. However, 
the second term, i.e., $c_3\tau(t)$, differs and is equal to $0$ when an 
$\omega_0$
exists such that $u = L\omega$, i.e., the signal generator can generate the input.
However, if no such $\omega_0$ exists, then the $\omega_0$ that results in the smallest~$\tau$ can be taken as in~\eqref{eq:choice_w0}.

Theorem~\ref{thm:observer_convergence_generic} certifies that the low-dimensional observer (\ref{eq:observer}) of user-defined order $\nu$ can reconstruct the system full state~$x$ with exponential convergence for specific classes of inputs characterized through the $\nu$ interpolation points in $\sigma(S)$.
This provides insight into the design of the observer as there is a trade-off between the size of the class of inputs for which~$\tau = 0$ and the number of interpolation points $\nu$.

\subsection{Systematic design of stable observer dynamics}
\label{S:MR:design}
This section presents a systematic way of designing the observer gain $K$ and the reduced-order matrix $G$ to ensure that $\Xi$ is Hurwitz as required by Theorem~\ref{thm:observer_convergence_generic}. 
The next result presents conditions under which $\Xi$ can be made Hurwitz by designing $G$ and $K$.

\begin{theorem}\label{thm:GK_selection}
    There exist $G,K\in\RR^{\nu}$  such that $\Xi$ in~\eqref{eq:error_dyn} is Hurwitz if and only if $(S,\begin{bmatrix} L^\top & (C \Pi)^\top \end{bmatrix}^\top)$ is detectable. \hfill$\Box$
\end{theorem}
\ifpreprint{
\begin{proof}
Under SA1, the matrix $A$ is Hurwitz.
Then, the matrix $\Xi$ is Hurwitz if and only if $S-GL-KC\Pi$ is Hurwitz thanks to its block diagonal structure as mentioned in Section \ref{S:MR:A}. 
The result of the theorem then trivially follows from a detectability argument.
\end{proof}
}\fi

In some cases, $G$ may be given as a result of a model reduction step first, or $K$ may already be known.
In such cases, the conditions can be presented as follows.
\begin{corollary}
    The following statements hold.
    \begin{itemize}
        \item Given a matrix $G \in \RR^{\nu}$, there exists $K\in\RR^{\nu}$ such that the matrix $\Xi$ in~\eqref{eq:error_dyn} is Hurwitz if and only if the pair $(S-GL,C\Pi)$ is detectable.
        \item Given a matrix $K \in \RR^{\nu}$, there exists $G\in\RR^{\nu}$ such that the matrix $\Xi$ in~\eqref{eq:error_dyn} is Hurwitz if and only if the pair $(S-KC\Pi,L)$ is detectable. \hfill $\Box$
    \end{itemize}
\end{corollary}
\ifpreprint{
\begin{proof}
    The proof follows immediately from the proof of Theorem~\ref{thm:GK_selection}.
\end{proof}
}\fi

The specific choice of $K=0$
is treated next as another corollary of Theorem~\ref{thm:GK_selection}.
\begin{corollary}\label{cor:K0}
    Let $K=0$.
    Then, there exists a matrix $G$ such that $S-GL$ is Hurwitz.
    Moreover, $S-G L$ is Hurwitz for the choice $G = \left(\Pi^\top P \Pi\right)^{-1} \Pi^\top P B$
    with $P \succ 0 $ the solution to $A^\top P + P A = - Q$ for any choice of $Q\succ 0$. \hfill $\Box$
\end{corollary}
\ifpreprint{
\begin{proof}
The proof that there exists a matrix $G$ such that $S-GL$ is Hurwitz follows immediately from the observability property in SA2.
    
The fact that $G = \left(\Pi^\top P \Pi\right)^{-1} \Pi^\top P B$ ensures that $S-G L$ is Hurwitz is proved as follows. 
    First note that $A^\top P + P A = -Q$ has a positive definite solution $P$ for any positive definite matrix $Q$ since $A$ is Hurwitz by SA1.
    Since $P \succ 0$ and rank$(\Pi) = \nu$ (see Section \ref{S:mm}), the matrix $\Pi^\top P \Pi \succ 0$, hence $\left(\Pi^\top P \Pi\right)^{-1}$ exists. 
    Next, note that the inequality 
    \[\Pi^\top(A^\top P + P A)\Pi \prec -\Pi^\top Q \Pi \prec 0 \]
    holds true since $\Pi$ is a full-column rank matrix.
    Using $A\Pi = \Pi S - BL$ in $\Pi^\top P A \Pi$ results in
    \[(\Pi S - BL)^\top P\Pi + \Pi^\top P (\Pi S-BL)\prec 0. \]
    Finally, using $\Pi^\top P B = \Pi^\top P \Pi G$, we verify that
    \begin{align*}&(\Pi S - \Pi G L)^\top P\Pi + \Pi^\top P (\Pi S-\Pi G L) \\
    &\quad=(S - G L)^\top \Pi^\top P\Pi + \Pi^\top P \Pi (S-G L) \prec 0
    \end{align*}
    holds true.
    Since $\Pi^\top P \Pi$ is positive definite, $(S-GL)$ must be Hurwitz, which completes the proof.
\end{proof}
}\fi

The result of this corollary shows that, for the case $K=0$, there always exists a specific $G$ that renders $S-GL$ Hurwitz.

\section{Numerical example}
\label{sec:example}

\begin{figure}
    \centering
    \includegraphics[width=0.95\linewidth]{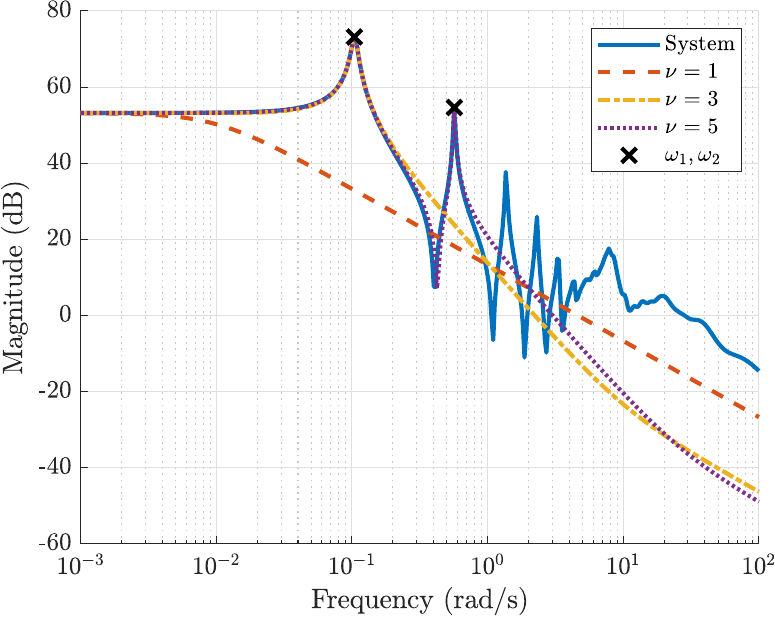}
    \caption{Bode magnitude plot of the clamped beam in Section~\ref{sec:example} together with three reduced-order models.\vspace{-6mm}}
    \label{fig:Ex_Bode}
\end{figure}

We illustrate the proposed approach on the benchmark model reduction problem of the clamped beam~\cite{antoulas2001survey}.
This model has $n=348$ states and satisfies SA1. 
Its input is the force applied at the free end, while its output is the resulting displacement,
with the corresponding Bode magnitude plot given in Figure~\ref{fig:Ex_Bode} (blue curve).

We design observers~\eqref{eq:ROM} with different state dimensions
namely for $\nu \in\{ 1,3,5\}$, which is a drastic reduction from the state dimension $n = 348$.
In this example, we define $\omega_1 \coloneqq 0.104$ and $\omega_2 \coloneqq 0.569$, and take $(S_\nu,L_\nu)$ as follows
\begin{align*}
    L_\nu &= \begin{bmatrix}
        1 & \ldots & 1
    \end{bmatrix}\in \RR^{1\times\nu},
    S_1 = 0, \\
    S_3 &= \textrm{blockdiag}(0,\Gamma(\omega_1)), S_5 = \textrm{blockdiag}(0,\Gamma(\omega_1),\Gamma(\omega_2)),
\end{align*}
where $\Gamma(\omega) \coloneqq \begin{bmatrix}
        0 & \omega \\
        -\omega & 0
    \end{bmatrix}, \omega \in \RR.$
It can be verified that SA2 holds for each considered value of $\nu$. 
The matrices $S_\nu$ correspond to the interpolation points $\sigma(S_1) = \{0\}$, $\sigma(S_3) = \{0,\pm j \omega_1\}$, and $\sigma(S_5) = \{0,\pm j \omega_1,\pm j \omega_2\}$.
The Bode magnitude plot of the corresponding reduced-order models with $G$ as in~Corollary~\ref{cor:K0} for $Q = I$ is depicted in Figure~\ref{fig:Ex_Bode}.
It can be observed that the corresponding frequency-response functions match at the corresponding interpolation points $\sigma(S_\nu)$.

\begin{figure*}[t]
    \centering
    \includegraphics[width=0.99\linewidth]{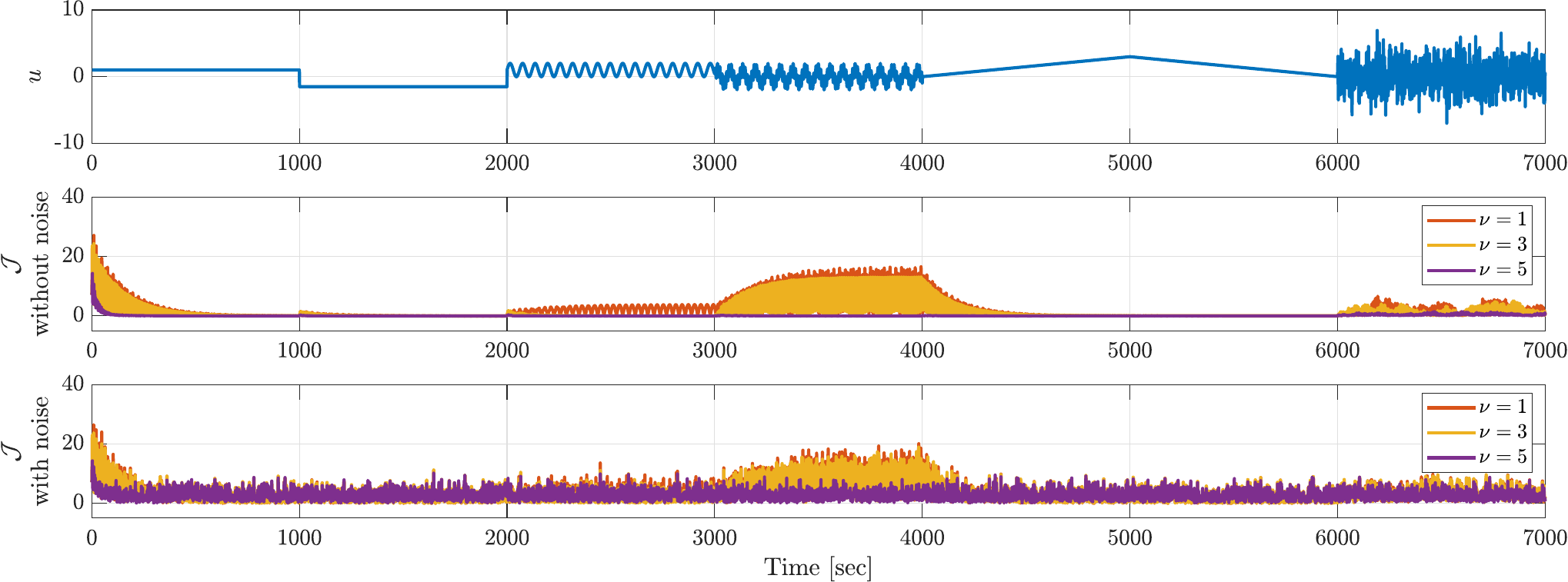}
    \caption{Simulation results for the clamped beam system. The top figure depicts the input signal, while the middle and bottom figures depict the performance measure~\eqref{eq:defJ} for the simulations without and with noise, respectively.\vspace{-3mm}}
    \label{fig:Ex_with_without_noise_combined}
\end{figure*}

To illustrate Theorem~\ref{thm:observer_convergence_generic}, we
take $ K = \begin{bmatrix}
        100 & \cdots & 100
    \end{bmatrix}^\top \in \RR^\nu,$
which ensures that $\Xi$ in~\eqref{eq:error_dyn} is Hurwitz for each considered dimension $\nu$.
The input $u$ is depicted in the top plot of Figure~\ref{fig:Ex_with_without_noise_combined},
where, for $t>6000$, $u(t)$ a zero-order-hold realization (with a sample time of 1 second) of a zero-mean white-noise sequence with a variance of 4.
The error at time $t \in [0,7T]$ with $T=1000$ is quantified using the measure $\mathcal{J}$ defined as 
\begin{equation}\label{eq:defJ}
\textstyle{
    \mathcal{J}(t) \coloneqq 100 \cdot \frac{\norm{e_{x\hat \xi}(t)}}{\norm{x}_{7T}},}
\end{equation}
i.e., the Euclidean norm of the error $e_{x\hat \xi}$ at time $t$ normalized by the $\mathcal{L}_\infty$-norm of $x$ over the simulation total time.
By this measure, $\mathcal{J} = 0$ means that the state is exactly reconstructed.

The middle plot of Figure~\ref{fig:Ex_with_without_noise_combined} depicts 
$\mathcal{J}$.
It can be concluded that all low-dimensional observers converge exponentially fast to the system state for constant inputs, i.e., up to $t = 2T$.
This is expected, as $0$ is an interpolation point in all observer designs.
Next, for $2T < t \le 3T$, the input is a sine wave with frequency~$\omega_1$.
As this frequency is matched in both the $\nu=3$ and $\nu=5$ case, these observers provide again convergence to zero error.
However, the observer corresponding to $\nu =1$, which does not match this frequency, yields a periodic error with frequency $\omega_1$.
Next, for $3T < t \le 4T$, the input is constituted of two sine waves with the frequencies $\omega_1$ and $\omega_2$.
It can be seen that only the observer corresponding to $\nu=5$ yields zero error.
The other two observers give a non-zero error.
Next, from $4T < t \le 6T$, the input is an increasing and decreasing ramp.
Since this ramp is sufficiently slow, all observers achieve an error close to zero.
Finally, when the input is a white-noise sequence, all errors are non-zero.
In conclusion, this example demonstrates the role of the interpolation points in the estimation error of the observers.
In particular, it is shown that full-state estimation can be achieved for certain classes of inputs.

Consider now the case in which the system output~$y$ is polluted with a zero-order-hold realization of a zero-mean white noise sequence with a variance that corresponds to a signal-to-noise ratio (SNR) of\footnote{The SNR is defined as $
    \text{SNR} = 10 \log_{10} \left( \frac{(y-\mu)^\top (y-\mu)}{\eta^\top \eta}\right)$, 
where $y$ represent the sequence of $\{y(t_k)\}_{k=0}^N$ at samples $t_k$ and $\mu$ is the mean of this sequence, and where $\eta$ is the white-noise sequence $\{\eta(t_k)\}_{k=0}^N$.} 20 dB.
The bottom plot of Figure~\ref{fig:Ex_with_without_noise_combined} shows the performance measure~$\mathcal{J}$, from which we conclude that the estimation error does not converge to exactly zero, but remains close to zero.
This error is consistent for all observers, except at the interval $3T < t \le 4T$, where the observers corresponding to $\nu=1$ and $\nu=3$ exhibit an error similar to the case without noise.

\section{Conclusions}\label{sect:conclusions}
This article presents a low-dimensional observer design approach for single-input single-output stable linear time-invariant systems. We have exploited recent advances in model order reduction for this purpose \cite{astolfi2010model}. In particular, the approach consists in first deriving a reduced-order model of the plant dynamics for a given class of inputs. 
An observer is then designed based on this reduced-order model, which generates estimates that are used to reconstruct the original full state of the plant model.
The results are demonstrated in a simulation on a benchmark model reduction problem.

In future work, we will extend the presented results to systems with possibly unstable state matrices and output nonlinearities thereby opening the door to the application to, e.g., lithium-ion battery models like those in \cite{blondel-et-al-tcst2018,khalil-et-al-tcst24,Dey_et_al_ACC_2014}. 

\ifpreprint{
\appendix

\section*{Proof of Proposition~\ref{proposition:observer_convergence_generic}}\label{appendix_A}

Let
$e=(e_{x\omega},  e_{\hat\xi\omega})\in \RR^{n}\times\RR^{\nu}$ and $Q \succ 0$.
As $\Xi$ is a Hurwitz matrix by assumption, there exists a unique $P \succ 0$ such that $P\Xi+\Xi^\top P=-Q$. 
We consider the Lyapunov function candidate $V(e):=e^\top P e$ for any $e \in \RR^{n+\nu}$. 
In view of \eqref{eq:error_dyn}, we have
\begin{multline}\label{eq:nabla_V}
        \langle \nabla V(e),\Xi e + \Psi(u-L\omega) \rangle \\ = e^\top (P\Xi+\Xi^\top P)e+2e^\top P\Psi(u-L\omega).
\end{multline}
By definition of $P$ and the Cauchy-Schwarz inequality,
\begin{multline}
        \langle \nabla V(e),\Xi e + \Psi(u-L\omega) \rangle \\ \leq  -e^\top Q e+2 \|e\| \|P\Psi\| \|u-L\omega\|.
\end{multline}
Using the fact that for any $p_1$, $p_2 \in \RR_{\ge0}$ and $\eta \in \RR_{>0}$, $2p_1p_2 \leq \tfrac{\eta}{2}p_1^2 + \tfrac{2}{\eta} p_2^2$, by taking $p_1=\|e\|$, $p_2=\|P\Psi\|\|u-L\omega\|$, and $\eta=\lambda_{\min}(Q)$, we obtain
\begin{align}
        \langle \nabla & V(e) , \Xi e + \Psi(u-L\omega) \rangle \notag
        \\&\leq  -\lambda_{\min}(Q) \|e \|^2 + \tfrac{\lambda_{\min}(Q)}{2}\|e \|^2  + \tfrac{2\|P\Psi\|^2}{\lambda_{\min}(Q)}\|u-L\omega\|^2 \notag
        \\ &\leq -\tfrac{\lambda_{\min}(Q)}{2}\|e \|^2+ \tfrac{2\|P\Psi\|^2}{\lambda_{\min}(Q)}\|u-L\omega\|^2. \label{eq:nabla_V_bound}
\end{align}
Using \eqref{eq:nabla_V_bound} and $\lambda_{\min}(P) \|e\|^2 \leq V(e) \leq \lambda_{\max}(P) \|e\|^2$, we derive \eqref{eq:ISS_generic_inputs} with $c_1:=\|\Phi\|\sqrt{\tfrac{\lambda_{\max}(P)}{\lambda_{\min}(P)}}$, $c_2:=\tfrac{\lambda_{\min}(Q)}{4\lambda_{\max}(P)}$, $c_3:=\tfrac{2 \|\Phi\| \|P\Psi\|}{\lambda_{\min}(Q)}\sqrt{\tfrac{\lambda_{\max}(P)}{\lambda_{\min}(P)}}$, and $\Phi$ the matrix in \eqref{eq:error_dyn} by following similar lines as in the proof of \cite[Theorem 4.10]{khalil_NLsystems_2002}.

 \medskip

}\fi


\balance
\bibliographystyle{ieeetr}
\bibliography{MyBIB,BIB_general}             

\end{document}